\documentclass[final,3p,times,twocolumn]{elsarticle}

\usepackage{graphicx}
\usepackage{amssymb}
\usepackage[dvips]{color}

\begin{document}

\begin{frontmatter}

\title{
Search for Heavy Neutrinos in  $\pi \rightarrow \mu \nu$ Decay
}

\author[labelmx]{A.~Aguilar-Arevalo}
\author[labelosaka]{M.~Aoki}
\author[labelvpi]{M.~Blecher}
\author[labelg]{D.I.~Britton}
\author[labelubc,labelnewdvm]{D.~vom Bruch}
\author[labelubc,label1]{D.A.~Bryman}
\author[labelcn]{S.~Chen}
\author[labelasu]{J.~Comfort}
\author[label1,labelnewld]{L.~Doria}
\author[labelubc]{S.~Cuen-Rochin}
\author[label1]{P.~Gumplinger}
\author[labelnubc,label1]{A.~Hussein}
\author[labelkek]{Y.~Igarashi}
\author[labelosaka,labelnewsi]{S.~Ito
\corref{cor2}}
\cortext[cor2]{Corresponding author: E-mail:s-ito@okayama-u.ac.jp}
\author[labelbnl]{S.H.~Kettell}
\author[label1]{L.~Kurchaninov}
\author[labelbnl]{L.S.~Littenberg}
\author[labelubc,labelnewcm]{C.~Malbrunot}
\author[label1]{R.E.~Mischke}
\author[label1]{T.~Numao
\corref{cor1}}
\cortext[cor1]{Corresponding author: E-mail:toshio@triumf.ca}
\author[labelg]{D.~Protopopescu}
\author[label1]{A.~Sher}
\author[labelubc,labelnewts]{T.~Sullivan}
\author[label1]{D.~Vavilov}

\address[labelmx]{Instituto de Ciencias Nucleares, Universidad Nacional Aut\'onoma de Mexico, D.F. 04510 M\'exico}
\address[labelosaka]{Graduate School of Science,
Osaka University, Toyonaka, Osaka 560-0043, Japan}
\address[labelvpi]{Physics Department, Virginia Tech.,
Blacksburg, Virginia, 24061, USA}
\address[labelg]{School of Physics and Astronomy, University of Glasgow, Glasgow, G12~8QQ, UK}
\address[labelubc]{Department of Physics and Astronomy,
University of British Columbia,
Vancouver, British Columbia, V6T 1Z1, Canada}
\address[label1]{TRIUMF, 4004 Wesbrook Mall, Vancouver,
 British Columbia, V6T 2A3, Canada}
\address[labelcn]{Department of Engineering Physics, Tsinghua University, Beijing, 100084, China}
\address[labelasu]{Physics Department, Arizona State University, Tempe, Arizona, 85287, USA}
\address[labelnubc]{University of Northern British Columbia,
Prince George, British Columbia, V2N 4Z9, Canada}
\address[labelkek]{KEK, 1-1 Oho, Tsukuba-shi, Ibaraki 305-0801, Japan}
\address[labelbnl]{Brookhaven National Laboratory, Upton, New York, 11973-5000, USA}
\address[labelnewdvb]{Present address: 
LPNHE, Sorbonne Universit\'{e}, Paris Diderot Sorbonne Paris Cit\'{e}, CNRS/IN2P3, Paris, France}
\address[labelnewld]{present address:PRISMA Cluster of Excellence and Institut f\"{u}r Kernphysik
Johannes Gutenberg-Universit\"{a}t Mainz, D 55128, Germany}
\address[labelnewsi]{Present address: Faculty of Science, Okayama University, 3-1-1 Tsushimanaka, Kita-ku,
Okayama, Japan 700-8530}
\address[labelnewcm]{Present address: CERN, 1211 Geneva 21, Switzerland}
\address[labelnewts]{present address: Department of Physics, Queen's University, Kingstone, Ontario, K7L 3N6, Canada}


\begin{abstract}

Heavy neutrinos were sought in pion decays $\pi^+ \rightarrow \mu^+ \nu$
by examining the observed muon energy spectrum for extra peaks
in addition to the expected peak for a massless neutrino.
No evidence for heavy neutrinos was observed.
Upper limits were set on the neutrino mixing matrix $|U_{\mu i}|^2$ 
in the neutrino mass region of 15.7--33.8 MeV/c$^2$, improving on previous results
by an order of magnitude.
\\

\end{abstract}


\date{\today}

\end{frontmatter}

\section{Introduction}
Neutrino oscillations indicate that at least two of the known neutrinos 
are massive, requiring the original Standard Model to be updated. 
The existence of additional heavy (mostly sterile) neutrinos  is still 
an open question. 
A wide range of motivations \cite{drews} for heavy neutrinos come from considerations of
baryogenesis \cite{canetti},  
large scale structure formation \cite{viel}, big bang nucleosynthesis \cite{ruch}, 
dynamical symmetry breaking \cite{appel}, and other effects \cite{gelmini,batell}. 
In general, the mass scale of heavy neutrinos is unknown opening up many possibilities 
and potential observables  in particle physics, astrophysics and cosmology. 
In the Neutrino Minimal Standard Model ($\nu$MSM) three sterile neutrinos 
are introduced which include a stable  dark matter candidate \cite{nmsm}.
MeV neutrinos could be accommodated in the $\nu$MSM  to
explain the $^7$Li abundance \cite{m10mev} or, with the addition of a MeV scalar, obtain 
consistent results for anomalies in  neutrino experiments \cite{model311}.

Massive neutrinos mixing with the muon neutrino
in the mass region 1--400 MeV/c$^2$
have been sought by accelerator-based experiments studying pion \cite{abela,dabtn,daum,psibeam} 
and kaon decays \cite{kek,e949,na62}.
The ratio of the $\pi^+ \rightarrow \mu^+ \nu_H$ decay rate to the normal 
$\pi^+ \rightarrow \mu^+ \nu$ decay rate can be written as

\begin{equation}
\frac{\Gamma(\pi^+ \rightarrow \mu^+ \nu_H)}{\Gamma(\pi^+ \rightarrow \mu^+ \nu)}
= |U_{\mu i}|^2 \overline{\rho}(m_H)
\end{equation}

\noindent
for $|U_{\mu i}|^2<<1$, 
where $m_H$ is the mass of the heavy neutrino $\nu_H$, $\overline{\rho}(m_H)$ accounts for the phase space and
helicity suppression factors (normalized to the zero-mass neutrino case) \cite{shrock}, 
and $U_{\mu i}$ is the neutrino mixing parameter
for the weak muon-neutrino eigenstate $\nu_{\mu}$ and the mass eigenstate $\nu_i$.

\begin{figure}[htb]
\centering
\includegraphics*[width=8.6cm]{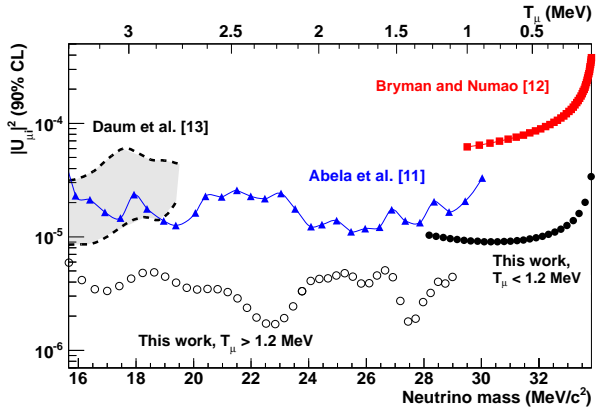}
\caption{
Summary of 90\% CL upper limits on $|U_{\mu i}|^2$ vs $m_H$.
The upper scale shows muon kinetic energy.
The blue line and red dots indicate the previous limits
by \cite{abela} and \cite{dabtn}, respectively.
Upper limits by \cite{daum}
are located in the shaded region between the two dashed lines.
The black closed-circles are the results of the present work
for the region $0<T_{\mu}<1.2$ MeV and the black open circles
for the region $1.1<T_{\mu}<3.3$ MeV.
}
\label{summary}
\end{figure}

Abela $et~al.$ \cite{abela} stopped pions in an active detector to measure the kinetic energy of muons
from the decay $\pi^+ \rightarrow \mu^+ \nu$
(kinetic energy $T_{\mu}$ = 4.1 MeV for a massless neutrino), and searched for low-energy
peaks due to massive neutrinos. Upper limits on $|U_{\mu i}|^2$ for the 5--30 MeV/c$^2$
region were set at a level of $10^{-2}-10^{-5}$; the experiment was limited by accidental background.
Massive neutrinos in the mass region 30--33.9 MeV/c$^2$ were also sought 
in pion decay by Bryman and Numao \cite{dabtn},
providing upper limits of $10^{-3}-10^{-4}$;
this search was limited by background coming from pion decays in flight ($\pi$DIF).
Daum $et~al.$ \cite{daum} employed a similar technique but with excellent energy resolution
using a Germanium detector;
the search region was 1-20 MeV/c$^2$ with a sensitivity level of $10^{-5}$,
which was limited by statistics.
A dedicated magnetic spectrometer experiment \cite{psibeam} searched 
for a 33.9 MeV/c$^2$ neutrino, providing a limit  of a 10$^{-8}$ 
level on  $|U_{\mu i}|^2$.
Figure \ref{summary} shows a summary of the present status of the search for massive neutrinos
at 90\% confidence level (CL)
in the mass region 15.7--33.8 MeV/c$^2$.

The present search for massive neutrinos was based on measurement of the muon kinetic energy  
in $\pi^+ \rightarrow \mu^+ \nu$ decays at rest.
When a pion is stopped in a thick plastic scintillator all decay vertices are contained;
the first signal is from the kinetic energy of the incident beam pion, 
the second from the muon in $\pi^+ \rightarrow \mu^+ \nu$ decay,
and the third from the positron in $\mu^+ \rightarrow \mbox{e}^+ \nu \overline{\nu}$ decay
($\pi^+ \rightarrow \mu^+ \rightarrow {\rm e}^+$ decay) which usually escapes in the case
considered here.
In the present study,
the leading backgrounds arising from accidental particles and $\pi$DIF
were suppressed using waveform and tracking information to identify extra
activity in the detector system, 
and the statistics were improved by an order of magnitude
over previous experiments.
\\

\section{Experiment}

\begin{figure}[htb]
\centering
\includegraphics*[width=7.7cm]{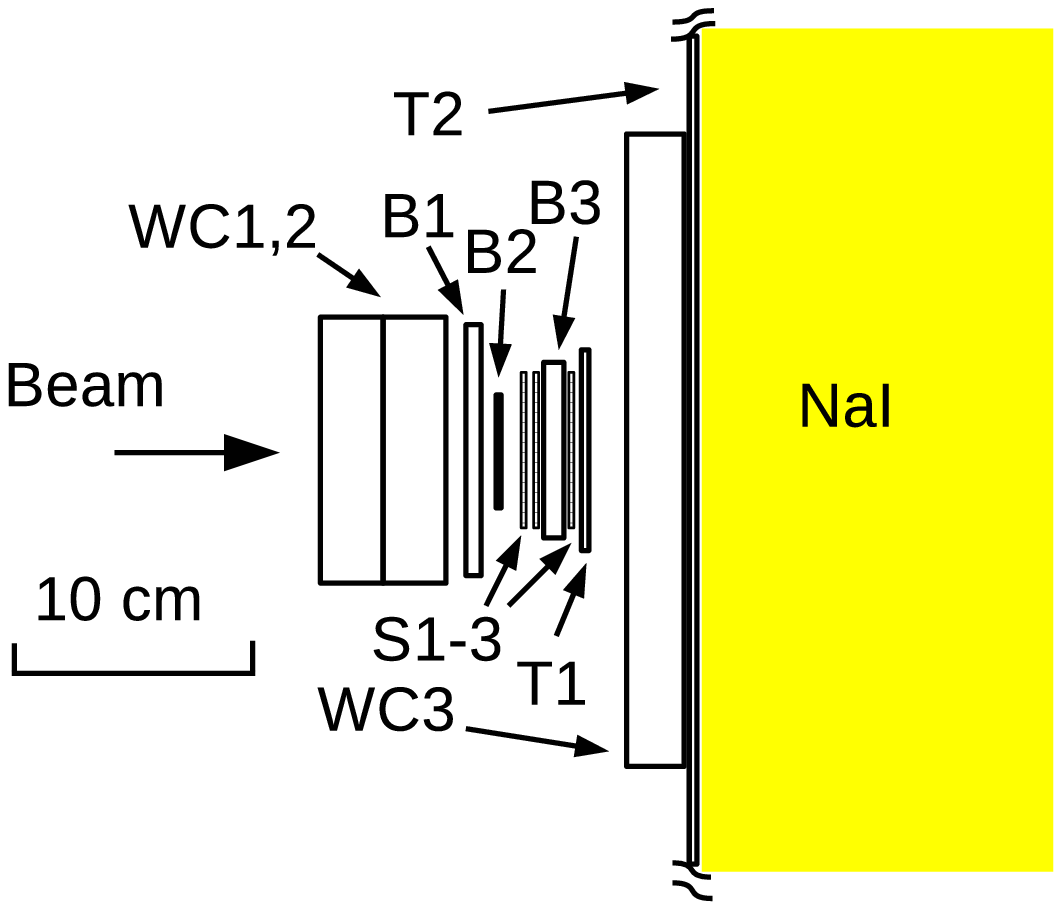}
\caption{Cross-sectional view of the PIENU detector.
CsI detectors surrounding the NaI crystal are not shown here.
}
\label{detector}
\end{figure}

The present search was carried out as a part of the PIENU experiment \cite{pienu},
a measurement
of the branching ratio $\Gamma (\pi^+ \rightarrow \mbox{e}^+ \nu (\gamma)) /
\Gamma (\pi^+ \rightarrow \mu^+ \nu (\gamma))$ using pion decays at rest,
where ($\gamma$) indicates inclusion of radiative processes.

Figure \ref{detector} shows a schematic view of
the PIENU apparatus \cite{nimpaper}.
A 75-MeV/$c$ $\pi^+$ beam from the TRIUMF M13 channel \cite{m13}
was degraded by two thin plastic scintillators B1 and B2
and stopped in an 8-mm thick plastic scintillator target
(B3) at a rate of $5 \times 10^4$ $\pi^+$/s.
The pion stopping-depth distribution in B3 was centered with 
the RMS width of 0.7 mm.
Pion tracking  was provided by wire chambers (WC1 and WC2)
 at the exit of the beam line,
and two X-Y sets of single-sided 0.3-mm thick planes
of silicon strip detectors, 
S1 and S2, placed immediately
upstream of B3.
Positrons from $\mu^+ \rightarrow {\rm e}^+ \nu \overline{\nu}$
decays were measured by another X-Y set of silicon strip detectors (S3), 
two thin plastic scintillators (T1 and T2),
a set of wire chambers (WC3),
and a calorimeter, consisting of 
a 48-cm (dia.) $\times$ 48-cm (length) single-crystal
NaI(T$\ell$) detector \cite{bnlbina} surrounded by
two concentric layers of pure CsI crystals \cite{e949csi}.

A positron signal from muon decay (defined by a T1-T2 coincidence)
in a time window  between --300 and 540 ns with respect to the incoming pion
 was the basis of the main trigger with a rate of $3 \times 10^3$ s$^{-1}$.
This was prescaled by a factor of 16 to form an unbiased trigger 
(Prescaled trigger).
Other triggers included a positron beam trigger for calibration
and ones for enhancing rare $\pi^+ \rightarrow \mbox{e}^+ \nu$ decays.
The typical trigger rate (including calibration triggers) was  600 s$^{-1}$.
The number of Prescaled-trigger events used in this analysis was $4 \times 10^9$.
\\

\section{Data processing and event selection}

The charge and time of each scintillator pulse were extracted from the waveform
recorded by a 500-MHz digitizer \cite{copper}.
The charge for each pulse was obtained by integrating the pulse from --20 to 20 ns 
around the local maximum for plastic scintillators  B1--3 and T1--2.
Also, there were several integration windows for each detector system to accommodate
specific purposes.
Each readout channel was calibrated to the expected energy loss
for 75-MeV/c beam particles.
The nonlinear response of plastic scintillator was studied in a Monte Carlo
simulation (MC) \cite{geant}
and consistency was confirmed within 0.1 MeV.

Events originating from pions were selected 
based on their energy loss
in B1 and B2,
and the presence of a B3 hit.
Events with extra activity other than that of the $\pi^+ \rightarrow \mu^+ 
\rightarrow {\rm e}^+$
decay sequence in B1, B2, B3, T1 and T2
within the time region of --6.4 to 1.4 $\mu$s with respect to the
pion stop were rejected.
This significantly reduced the accidental background.

In order to ensure that an event was due to $\pi^+ \rightarrow \mu^+ \rightarrow {\rm e}^+$
decay, the calorimeter energy was required to be $<$55 MeV 
and a loose geometrical positron acceptance cut 
(the radius of the hit position at WC3 to be $<$80 mm)
was applied.
The number of $\pi^+ \rightarrow \mu^+ \rightarrow {\rm e}^+$ events was $5 \times 10^8$.

A second pulse in B3 due to the muon kinetic energy
needed to be identified in the search.
However,
when the muon energy involving $\nu_H$ was below 1.2 MeV 
the pulse detection logic could not efficiently identify the pulse.
The search was, therefore, divided into two muon energy regions, above and below 1.2 MeV.
In the region below 1.2 MeV the integrated energy up to 600 ns, containing the
$\pi^+ \rightarrow \mu^+ \rightarrow {\rm e}^+$ decay sequence, was examined 
for the muon energy contribution,
while in the region above 1.2 MeV the energy of a cleanly separated second pulse
was used.
In each analysis, all $\pi^+ \rightarrow \mu^+ \rightarrow {\rm e}^+$ decay events were
examined.
\\

\section{Analysis of the region below 1.2 MeV}

Since the B3 energy observed with the wide time window included the pion and positron contributions
in addition to the kinetic energy of the muon,
these energies were first subtracted.

The contribution from the positron pulse, which resulted in a wider energy
distribution due to the path length variation of the positron in B3, was subtracted
using a well separated positron pulse in B3.
In order not to affect the $\pi^+ \rightarrow \mu^+ \nu$ decay measurement,
late positron time
was required ($300 < t_{\rm T1}-t_{\rm B1} < 500$ ns).

\begin{figure}[htb]
\centering
\includegraphics*[width=8.5cm]{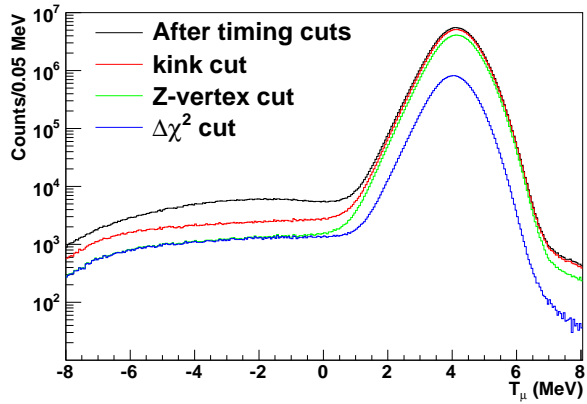}
\caption{
Energy spectra of muons 
for the $T_{\mu} < 1.2$ MeV analysis
after subtraction of pion and positron energies
from the total energy, with several cuts;
black indicates the spectrum after the positron time cut, red after the angle cut, 
green after the Z vertex cut
and blue after the $\Delta \chi^2$ cut.
}
\label{energy2}
\end{figure}

In order to correct for the energy-loss variation of pions in the upstream detectors, 
the energies of the other beam counters (B1, B2, S1 and S2) were added for this search.
Then, the contribution of the visible pion kinetic energy ($\sim$17 MeV) to the total energy was subtracted
by shifting the spectrum to align the peak at 4.1 MeV.
The black histogram in Fig. \ref{energy2} shows the spectrum of the muon energy after the positron
and pion energies were subtracted from the total energy.
\\


The main background for this search is $\pi$DIF occurring very near or in B3,
in which the observed ``pion'' energy can be less than the incident pion beam energy.
The low-energy tail of the 4.1 MeV peak extending to $T_{\mu} = 0$ MeV
was another background.
The contribution from pion radiative decay $\pi \rightarrow \mu \nu \gamma$ was negligible.

The $\pi$DIF background was suppressed by two cuts \cite{ournu}.
The angle between the track vectors measured by WC1-2 (upstream) and S1-2 (near B3) 
provided a measure of a kink in the pion track
due to $\pi$DIF. Limiting the kink angle  suppressed the $\pi$DIF 
background as shown by the red histogram in Fig. \ref{energy2}.
The muon decay vertex in the beam direction (Z-vertex)
obtained from the pion (WC1-2 and S1-2) and positron (S3 and WC3) tracks 
also allowed identification of $\pi$DIF in B3
as shown by the green histogram.
The low energy tail of the 4.1 MeV peak,
mainly due to cleanly separated $\pi^+ \rightarrow \mu^+ \rightarrow {\rm e}^+$ decays,
was suppressed by a consistency test
based on the difference of $\chi^2$ values
per degrees of freedon for two- and three-pulse fits.
In this $\Delta \chi^2$ cut, only three pulse events with the second pulse 
falling into $T_{2nd} > 2$ MeV were rejected, so that
the acceptance loss for the
signal (T$_{\mu} < 1.2$ MeV) was negligible.
After all cuts, the small but dominant background was still from $\pi$DIF in B3
as shown by the blue histogram in Fig. \ref{energy2}.
\\

\subsection{Fit}

There are two background shapes to be reproduced
at energies above and below the signal region (0--1.2 MeV):
the 4.1 MeV peak and the $\pi$DIF shape.
The slightly asymmetric shape of the 4.1 MeV peak was mostly due to the
kinetic energy distribution of the pions.
The energy spectrum of events rejected by the $\Delta \chi^2$ cut was used to represent the
background peak at 4.1 MeV.
Since the events rejected by the $\Delta \chi^2$ cut had a bias
due to the requirement of $>$2 MeV muon energy,
the peak position and resolution were slightly
different from those of the remaining events. The bias was compensated
by multiplying the background peak by a unit Gaussian function with its center shifted by 0.1 MeV.
The amplitude of the background peak 
and the resolution of the unit Gaussian were free parameters of the fit.
A quadratic function was used for the $\pi$DIF background.

\begin{figure}[htb]
\centering
\includegraphics*[width=8.5cm]{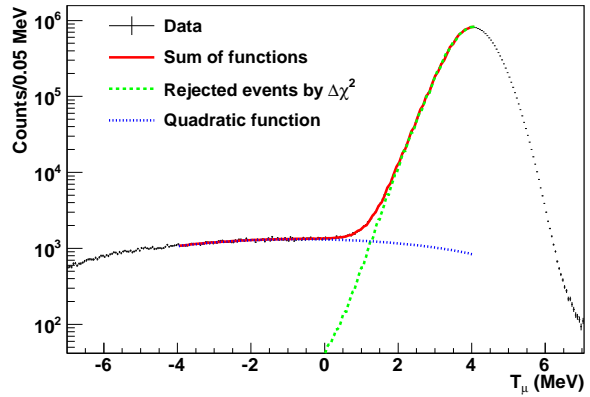}
\caption{
Fit results for the region $T_{\mu}<1.2$ MeV. The black histogram, 
the blue dot and green dashed lines indicate
data, quadratic background and the 4.1 peak, respectively.
The smearing was applied in the plot.
}
\label{fit2}
\end{figure}

Figure \ref{fit2} shows the fit using the two shapes described above.
The fitting window of --4.0 to 3.5 MeV resulted in $\chi^2 /ndof = 1.39$;
there is some deviation above 2 MeV 
due to a small remaining mismatch in the resolution and peak position.

The signal peak in the region $T_{\mu}<1.2$ MeV
was assumed to be a Gaussian with a resolution  offset.
The resolution was scaled
by the square root of the energy from that of the 4.1 MeV peak,
and the offset was obtained from the pion-only peak (to represent a peak at 0 MeV)
obtained from an 80-ns integration window with a late muon pulse.
The signal peak energy was varied from 0.0 MeV to 1.3 MeV with 0.05 MeV steps.

\begin{figure}[htb]
\centering
\includegraphics*[width=8.5cm]{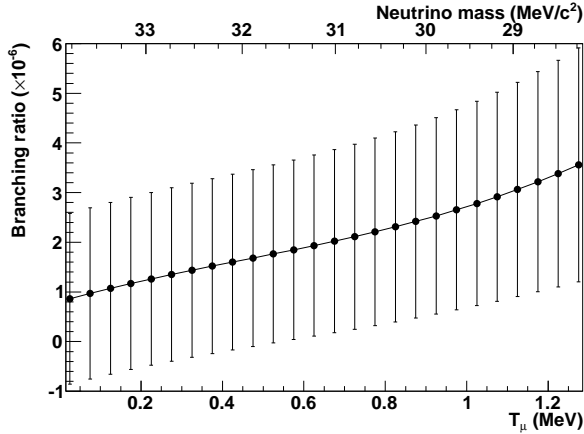}
\caption{
Normalized amplitudes (branching ratio) vs kinetic energy
with statistical errors ($\pm 1 \sigma$) after the Z-vertex acceptance correction.
The upper scale shows $m_H$.
Note: The step size is much smaller than the resolution.
}
\label{branchingratio2}
\end{figure}

Figure \ref{branchingratio2} shows the signal amplitudes, with
statistical errors, normalized to
that of the 4.1 MeV peak 
(without the two-pulse requirement) after the acceptance correction
of 1.06 for the Z-vertex cut; the distribution of the events in the 4.1 MeV peak was wider 
due to the longer muon range.
Since no significant signal beyond the statistical fluctuation
was observed,
90\% CL upper limits on the neutrino mixing parameter $|U_{\mu i}|^2$ were obtained
according to the Bayesian procedure assuming a positive peak amplitude
with a Gaussian probability distribution.
The full black circles for $0 < T_{\mu} <1.2$ MeV (top axis)
in  Fig. \ref{summary} show the result.
\\

\section{Analysis of the region above 1.2 MeV}

In this analysis 
the energy of the second pulse in the $\pi^+ \rightarrow
\mu^+ \rightarrow {\rm e}^+$ decay in B3 was used.
The largest background 
was accidental background
originating from decay of an
``old'' muon from an earlier pion or muon residing near B3
to a positron,
giving a second or third pulse;
some accidental background was already suppressed by the requirement of no additional
coincident activity in the detectors.
Radiative pion decay $\pi^+ \rightarrow \mu^+ \nu \gamma$ 
(branching fraction, $2 \times 10^{-4}$ \cite{radpi}) was about one tenth of the
accidental background in the previous experiment \cite{abela}.
\\

\begin{figure}[htb]
\centering
\includegraphics*[width=8.5cm]{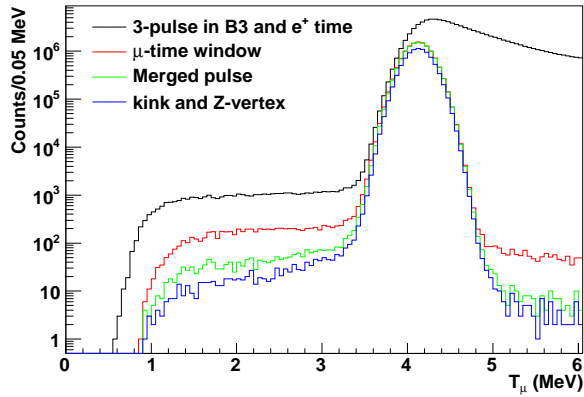}
\caption{
Energy spectra of the second pulse.
The black line is for positrons in the 200--500 ns window as the third pulse,
red with the time window cut of 80--150 ns for the second pulse, green with the merged-pulse cut in B3, 
and blue with the angle and  vertex cuts.
}
\label{energy1}
\end{figure}

Events with three separate pulses in B3 were selected in this analysis.
Figure \ref{energy1} shows energy spectra of the second pulse
with consecutive cuts described below.
In order to clearly separate the positron pulse from the muon pulse,
events with a late positron signal, firing T1 and 
the calorimeter in the positron time region $200 < t_{T1}-t_{B1} <500$ ns,
were selected.
The second pulse was required to be within 80--150 ns with respect to the first pulse (pion stop).
With this cut the energy variation due to the overshoot of the first pulse
was minimized to $<$0.16 MeV (a negligible level in terms of resolution).

The comparison of the pulse height and the integrated charge (--20 to 20 ns)
of the first pulse
identified early pion decays in which the muon pulse merges into the first pulse
and a positron pulse is recognized as the second.
The effect is shown by the green histogram in Fig. \ref{energy1}.
The kink angle and Z-vertex cuts reduced the contribution of $\pi$DIF.
The total number of events used for the search in the region $T_{\mu}>$1.2 MeV was $9 \times 10^6$.
After these cuts,
the radiative pion decay $\pi^+ \rightarrow \mu^+ \nu \gamma$ 
was the major source of background.
\\

\subsection{Fit}

The background shape  for radiative pion decay was generated
by MC \cite{geant} with the selection cuts,
and the amplitude was a free parameter in the fit.
A constant term (not in the final fit) was also added as a free parameter
to represent other potential backgrounds, but it was always consistent with zero.
The peak at 4.1 MeV was fitted with a Gaussian with the resolution,
amplitude and position as free parameters.
The acceptance losses due to the selection requirements 
were common to those of the peak at 4.1 MeV
except for the pulse-detection logic which depended on the
amplitude of the pulse (note the sharp drop around 1 MeV).
This acceptance was estimated by comparing
the observed positron energy spectrum with the MC positron spectrum which was not affected
by the effects of the detection logic.
Since the background events were also affected by the pulse detection logic,
the fitting functions of the background and the signal were multiplied by this 
acceptance function.

\begin{figure}[htb]
\centering
\includegraphics*[width=8.5cm]{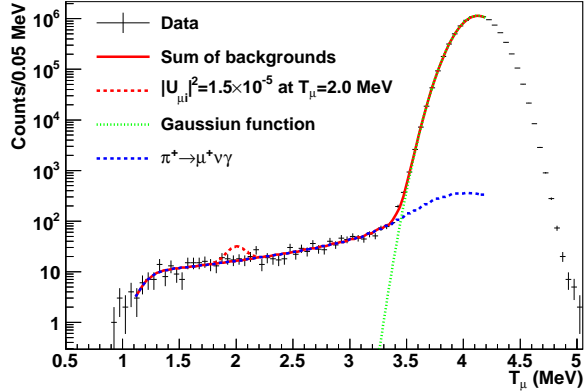}
\caption{
Data are shown by the black cross with statistical errors. The red lines are for the fit results
without (solid) and with (dashed) a signal at 2 MeV for $|U_{\mu i}|^2=1.5 \times 10^{-5}$.
The green dotted line is for the
4.1 MeV Gaussian peak and the blue dashed line for the radiative decay background.
}
\label{fit1}
\end{figure}

A fit in the energy region $1.1 < T_{\mu} < 4.2$ MeV with a no-signal assumption
provided a total $\chi^2 = 68$ with 58 degrees of freedom.
The largest deviations were in the region above 3.4 MeV.
Figure \ref{fit1} shows the data in black and the result of the background fit 
in the red solid line.
The small peak at 2 MeV (the red dashed line) indicates a hypothetical signal 
for $|U_{\mu i}|^2=1.5 \times 10^{-5}$.

\begin{figure}[htb]
\centering
\includegraphics*[width=8.5cm]{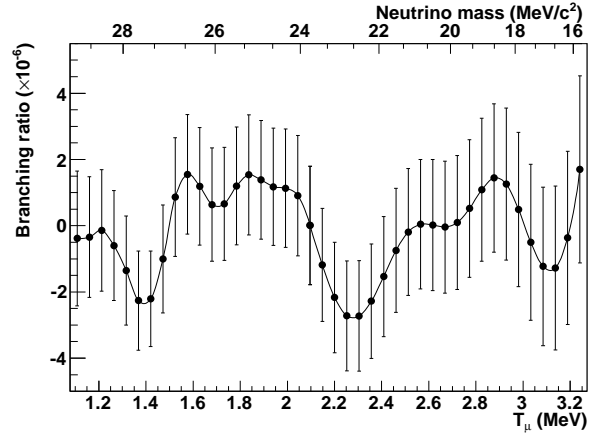}
\caption{
Normalized amplitudes (branching ratio) of the second pulse vs kinetic energy
with statistical errors ($\pm 1 \sigma$).
The upper scale shows $m_H$.
}
\label{branchingratio1}
\end{figure}

For the search for additional peaks,
the signal function, a Gaussian with the resolution scaled by the square root of the energy
from that of the 4.1 MeV peak,
was added to the fitting function.
The signal  peak was shifted in 0.05 MeV steps for the search.
The amplitude of the signal peak was an additional free parameter of the fit.
No significant excess above statistical fluctuations was observed.
The amplitudes of hypothetical peaks normalized to the 4.1 MeV peak, with the Z-vertex
acceptance correction, are shown in Fig. \ref{branchingratio1}.
The amplitudes in the region T$_{\mu}>$3.4 MeV  monotonically rise, reflecting
a small deviation from the peak function.
Since we did not see a significant excess in the signal amplitudes, 
a 90\% CL upper limit on the mixing matrix
$|U_{\mu i}|^2$ was set for each neutrino mass.
The black open circles for $1.1 <  T_{\mu} < 3.3$ MeV
in Fig. \ref{summary} show 90\% CL upper limits for $|U_{\mu i}|^2$.
\\

\section{Conclusion}

No evidence for massive neutrinos in $\pi^+ \rightarrow \mu^+ \nu$ decay was observed.
Upper limits on the neutrino mixing parameter $|U_{\mu i}|^2$
were obtained for the region
$0 < T_{\mu} < 3.3$ MeV, which corresponds to
$15.7 < m_H < 33.8$ MeV/c$^2$.
The improvement factors were approximately an order of magnitude
compared to the previous
searches \cite{abela,dabtn,daum}.\\

\section{Acknowledgments}

This work was supported
 by the Natural Sciences and Engineering Research Council
and TRIUMF through a contribution from the National Research Council of Canada,
and by the Research Fund for the Doctoral Program of Higher Education of China, 
by CONACYT doctoral fellowship from Mexico, and
by JSPS KAKENHI Grant numbers 18540274, 21340059, 24224006,
19K03888 in Japan.
We are grateful to Brookhaven National Laboratory for
the loan of the crystals, and to the TRIUMF operations, detector, 
electronics and DAQ groups for
their engineering and technical support.
\\

\end{document}